\pdfoutput=1
\documentclass{appolb}
\usepackage{hyperref}
\usepackage{epsfig}
\usepackage{amsmath}%for subequations
\usepackage{amssymb}%for symbols
\usepackage{subfigure}%for subnumbering inside a figure
\usepackage{color}
\newcommand{\bk}{{\bf{k}}}
\newcommand{\ba}{{\bf{a}}}

\begin{document}

\title{
%%%%%%%%%%%%%%%%%%
\vspace{-20mm}
\begin{flushright} \bf IFJPAN-IV-2010-2\\ \end{flushright}
\vspace{5mm}
%%%%%%%%%%%%%%%%%%
Properties of inclusive versus exclusive QCD evolution kernels.%
\thanks
{This work is supported by the EU grant MRTN-CT-2006-035505, 
 by the Polish Ministry of Science and Higher Education grant 
  No.\ 153/6.PR UE/2007/7, and by the Marie Curie research training network ``MCnet'' (contract number MRTN-CT-2006-035606.
\hfill \\
  Presented by A.~Kusina at the
{\em Cracow Epiphany Conference 2010 - physics in underground
laboratories and its connection with LHC}, January 5-8, 2010}
}%
\author{A. Kusina\\
 S. Jadach, M. Skrzypek and M. S\l{}awi\'nska
\address{Institute of Nuclear Physics PAN,\\
ul. Radzikowskiego 152, 31-342 Krak\'ow, Poland }}
\maketitle

\begin{abstract}
{\em Abstract:}
We investigate the role of the choice of the upper phase
space limit $Q$ in the Curci-Furmanski-Petronzio (CFP)
factorization scheme,
which exploits dimensional regularization $MS$ scheme.
We examine how the choice of $Q$ influences the evaluation of
the standard DGLAP (inclusive) evolution kernels,
gaining experience needed in the construction of
the {\em exclusive} Monte Carlo modelling of the NLO DGLAP evolution.
In particular, we uncover three types of
mechanisms which assure the {\em independence} on $Q$ of the inclusive
DGLAP kernels calculated in the CFP scheme. We use the examples 
of three types of the Feynman diagrams to illustrate our analysis.

\vspace{3mm}
\centerline{\em Submitted To Acta Physica Polonica B}
\end{abstract}

\PACS{12.38.-t, 12.38.Bx, 12.38.Cy}

\vspace{5mm}
%%%%%%%%%%%%%%%%%%%%%%%%%%%
\begin{flushleft}
\bf IFJPAN-IV-2010-2\\
\end{flushleft}
%%%%%%%%%%%%%%%%%%%%%%%%%%%

\newpage
%%%%%%%%%%%%%%%%%%%%%%%%%%%%%%%%%%%%%%%%%%%%%%%%%%%%%%%%%%%%%%%%%%%%%%%
%%%%%%%%%%%%%%%%%%%%%%%%%%%%%%%%%%%%%%%%%%%%%%%%%%%%%%%%%%%%%%%%%%%%%%%
\section{Introduction}
%%%%%%%%%%%%%%%%%%%%%%%%%%%%%%%%%%%%%%%%%%%%%%%%%%%%%%%%%%%%%%%%%%%%%%%
This study is part of an effort aiming at construction
of an {\em exclusive} Monte Carlo modeling of DGLAP~\cite{DGLAP}
evolution of the parton distribution functions (PDFs)
in the next-to-Leading-Order (NLO) approximation
using work of Curci-Furmanski-Pertonzio (CFP)~\cite{Curci:1980uw}
as a starting point.
Standard {\em inclusive} PDFs are defined within the
framework of the collinear factorization
theorems~\cite{Ellis:1978sf,Collins:1984kg,Bodwin:1984hc}.
The ongoing project of defining and implementing in the MC form
exclusive PDFs (ePDFs), see ref.~\cite{Jadach:2009gm,Jadach:2010ew},
sometimes also referred to as {\em fully unintegrated} PDFs
\cite{Hautmann:2009zz},
is based on the older formulation of
the collinear factorization of ref.~\cite{Ellis:1978sf}
reformulated later on by CFP~\cite{Curci:1980uw}.
The CFP work uses dimensional regularization in $MS$ scheme
and physical axial gauge.

The construction of ePDFs in the Monte Carlo form
requires defining and calculating new {\em exclusive} evolution kernels.
Moreover, it is critical to understand and analyse the
properties of exclusive kernels, especially
cancellations of the infrared singularities diagram by diagram
due to gauge invariance, see study in ref.~\cite{Slawinska:2009gn}.

In this contribution we comment on the issue of
the {\em independence} of
inclusive/exclusive kernels on the choice of the upper phase
space limit $Q$ in their evaluation based on the Feynman diagrams.
Of course, the independence of
the inclusive DGLAP evolution kernels in the CFP
may be regarded as obvious and trivial.
However, in the actual calculation of the kernel from the Feynman diagrams
the genuine mechanism which protects its independence on $Q$
looks rather mysterious and not obvious at all.
The choice of $Q$ turns out to be important in the analytical
evaluation of the NLO kernels in the CFP scheme, because it
determines quite rigidly the parametrization of the two-parton phase space.
In addition, in the construction of Monte Carlo model for ePDF
the same upper phase space limit variable $Q$ is
closely related  to the {\em evolution time variable}.
It is therefore quite interesting to have a closer look
into the above phenomena.

\begin{figure}[h!]
\begin{centering}
\subfigure[]{
\includegraphics[height=2cm]{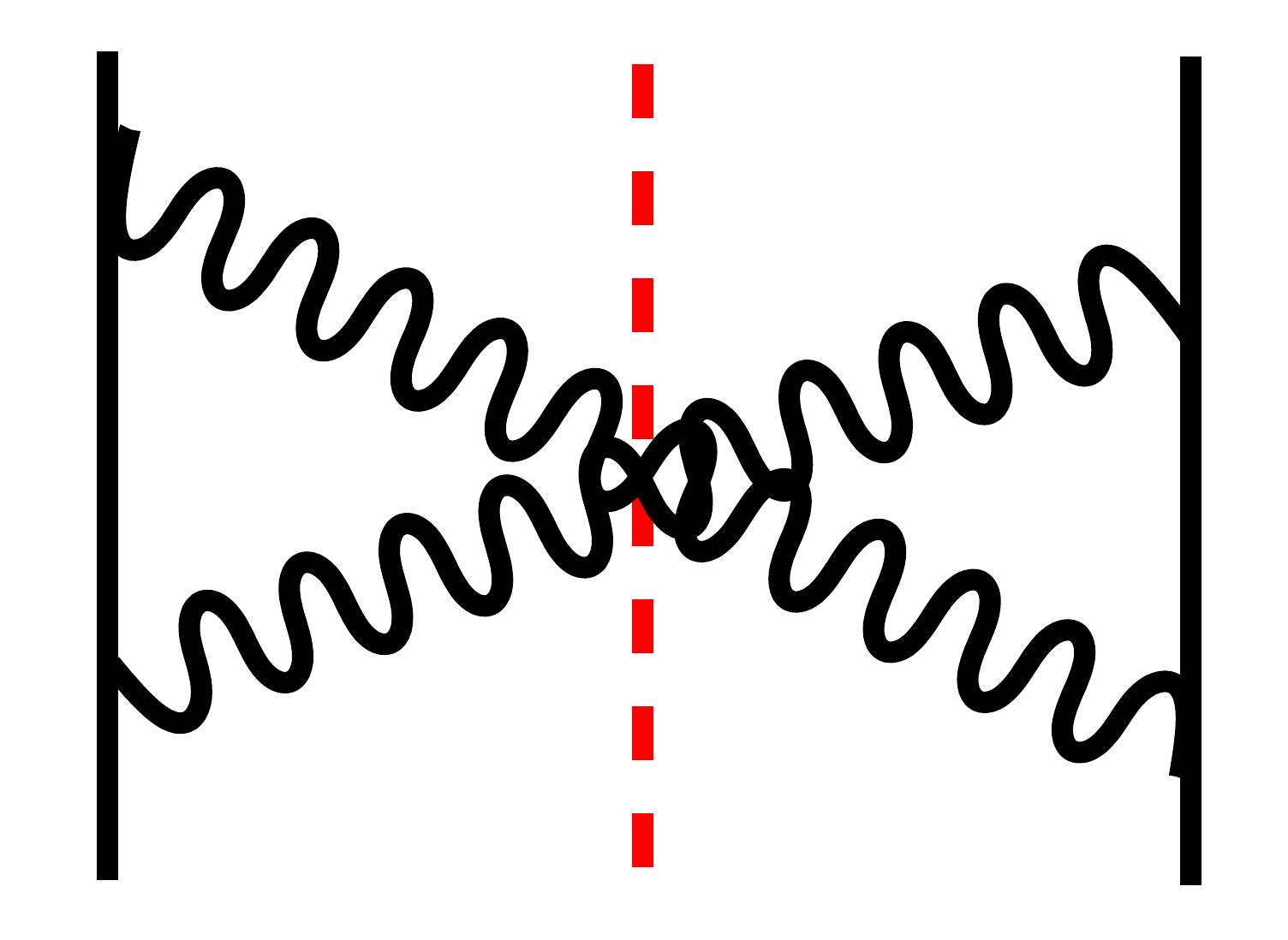}
  \label{subfig:Bx}}
\subfigure[]{
\includegraphics[height=2cm]{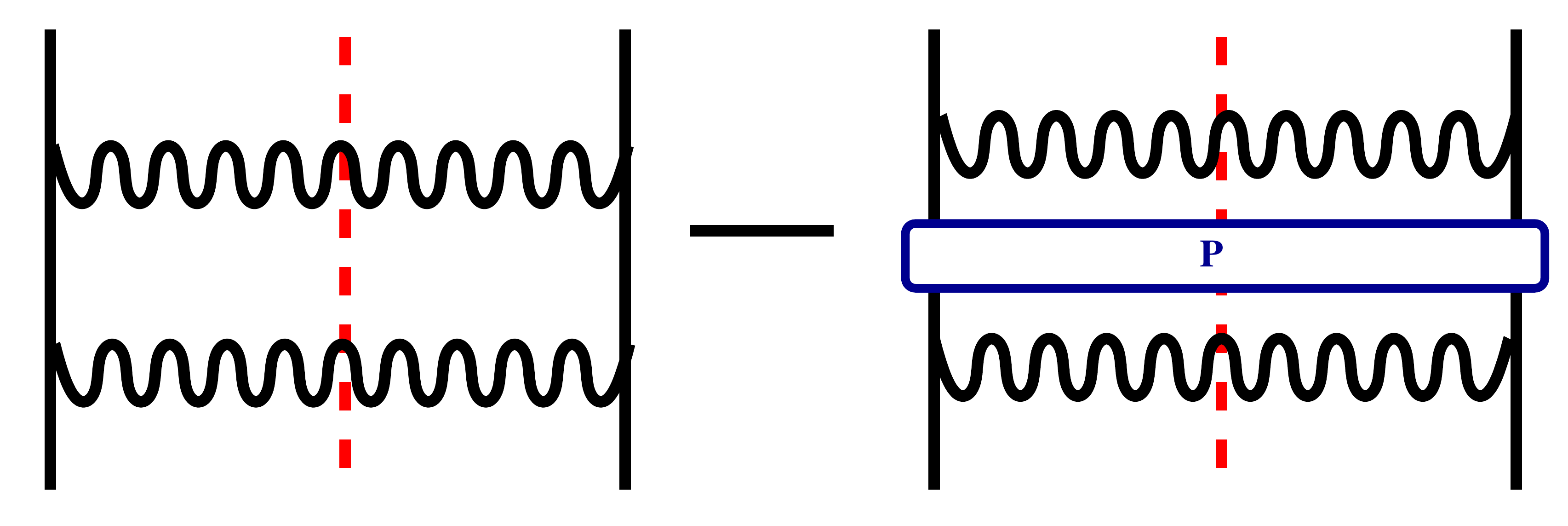}
 \label{subfig:Br-K}}
\subfigure[]{
\includegraphics[height=2cm]{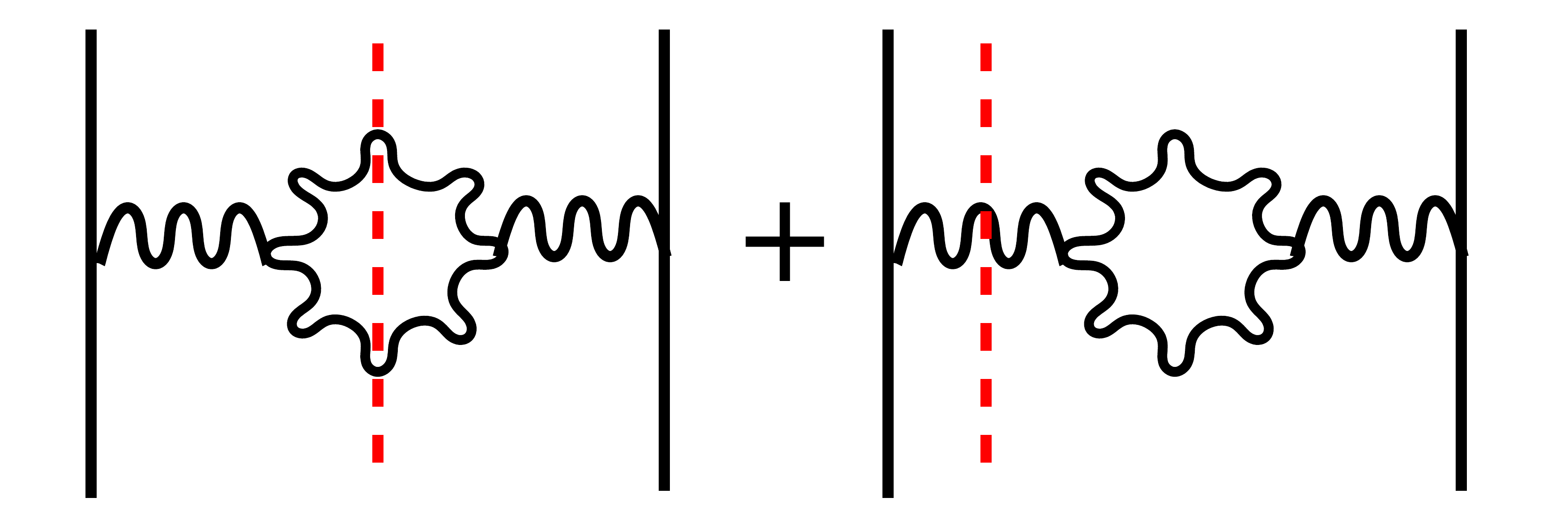}
 \label{subfig:Vg}}
\caption{Example Feynman diagrams contributing to NLO DGLAP kernel.}
\label{fig:graphs}
\end{centering}
\end{figure}

In the following 
we will show that there are three different mechanisms which assure
the {\em independence} of inclusive NLO DGLAP
evolution kernels on the upper phase space limit $Q$ in the CFP scheme.
We demonstrate each mechanism using an example of the
Feynman diagram contributing to NLO DGLAP kernel.
We shall use subset of diagrams shown in Figure~\ref{fig:graphs}.

%%%%%%%%%%%%%%%%%%%%%%%%%%%%%%%%%%%%%%%%%%%%%%%%%%%%%%%%%%%%%%%%%%%
%%%%%%%%%%%%%%%%%%%%%%%%%%%%%%%%%%%%%%%%%%%%%%%%%%%%%%%%%%%%%%%%%%%
\section{Notation}
%%%%%%%%%%%%%%%%%%%%%%%%%%%%%%%%%%%%%%%%%%%%%%%%%%%%%%%%%%%%%%%%%%%
\label{sec:bremss}

We consider two-gluon real emission diagrams.
For the four-momentum parametrization we use Sudakov variables:
\begin{equation}
k_i = \alpha_i p + \beta_i n + k_{i\perp}, \quad i = 1,2,
\end{equation}
with $p$ being the four-momentum of the incoming quark
and $n$ a light-cone vector. Four-vectors of two
emitted gluons are $k_1$ and $k_2$, with their
transverse parts being $k_{1\perp}$ and $k_{2\perp}$
respectively, and $k^2=(k_1+k_2)^2$ being their effective mass.
Since the emitted gluons are on mass shell and we are in the
massless theory, $\beta_i$ are fixed and equal to
$\beta_i=-\frac{k_{i\perp}^2}{2\alpha_i(pn)}
=\frac{\bk_{i\perp}^2}{2\alpha_i(pn)}$.
We will also use $q$ symbol for the off-shell momentum
$q=p-k=p-k_1-k_2$.

In the CFP scheme~\cite{Curci:1980uw}
the contribution of each Feynman diagram to the DGLAP kernel is
extracted  from the phase space integral:
\begin{equation}
P(x) = {\rm Res}_0\bigg( \int d\Phi\; \delta(1-x-\alpha_1-\alpha_2)
              \;\rho(k_1,k_2)\; \Theta(s(k_1,k_2)\le Q)\bigg),
\label{eq:gamma}
\end{equation}
where ${\rm Res_0}$ is the residue at $\epsilon=0$ 
(the coefficient in front of $\frac{1}{\epsilon}$ pole
in the dimensional regularization),
$\delta(1-x-\alpha_1-\alpha_2)$ is the definition of the Bjorken
variable, $\rho$ is a contribution from a Feynman diagram
(originating from $\gamma$-traces, etc.).
The element of  the two gluon phase space $d\Phi$ is given by:
\begin{equation}
  d\Phi = \frac{d^nk_1}{(2\pi)^n} 2\pi \delta^+(k_1^2)
          \frac{d^nk_2}{(2\pi)^n} 2\pi \delta^+(k_2^2).
\end{equation}
The theta function in equation \eqref{eq:gamma}
encloses the phase space from above
using dedicated kinematical variable $s(k_1,k_2)$.
The choice of  phase space enclosing, $s(k_1,k_2)$,
{\em determines} the choice of {\em evolution time variable}
in the construction of the MC implementation of ePDF.
There are many possible choices for $s(k_1,k_2)$ function,
here we will concentrate on two of them:
$s(k_1,k_2)=\max\{|\bk_{1\perp}|,|\bk_{2\perp}| \}$, which corresponds
to the transverse momentum evolution time and
$s(k_1,k_2)=\max\{|\ba_1|,|\ba_2| \}$, which corresponds to rapidity
related evolution time. 
Scalar quantity $a_i$ is defined as a modulus
of the vector variable:
\begin{equation}
  \ba_i=\frac{\bk_{i\perp}}{\alpha_i}
\end{equation}
and we call it {\em angular scale} variable.
It is related to rapidity via $y_i=\ln|\ba_i|$.
These two cases will be respectively referred to as phase space with
$k_\perp$-ordering and angular-ordering ($a$-ordering).
Other popular choices  of $s$-function
include total virtuality $\sqrt{-q^2}$
and maximum $k$-minus, $\max(k^-_1,k^-_2)$.

Since we will show calculations in both angular ordered and
$k_{\perp}$-ordered phase space we give the phase space
parametrization in both sets of variables (remembering that we
work in dimensional regularization with number of dimensions
equal to $n=4+2\epsilon$, $\epsilon>0$):
\begin{equation}
d\Phi_{k_{\perp}} = \frac{1}{4\mu^{4\epsilon}}
                    \frac{\Omega_{1+2\epsilon}}{(2\pi)^{6+4\epsilon}}
                    \frac{d\alpha_1}{\alpha_1}\frac{d\alpha_2}{\alpha_2}
                    d\Omega_{1+2\epsilon} dk_{1\perp}dk_{2\perp}
                    k_{1\perp}^{1+2\epsilon}k_{2\perp}^{1+2\epsilon},
\end{equation}
and
\begin{equation}
d\Phi_{a} = \frac{1}{4\mu^{4\epsilon}}
            \frac{\Omega_{1+2\epsilon}}{(2\pi)^{6+4\epsilon}}
            \frac{d\alpha_1}{\alpha_1}\frac{d\alpha_2}{\alpha_2}
            \alpha_1^{2+2\epsilon}\alpha_2^{2+2\epsilon}
            d\Omega_{1+2\epsilon} da_1da_2
            a_1^{1+2\epsilon} a_2^{1+2\epsilon}.
\end{equation}

%%%%%%%%%%%%%%%%%%%%%%%%%%%%%%%%%%%%%%%%%%%%%%%%%%%%%%%%%%%%%%%%%%%
%%%%%%%%%%%%%%%%%%%%%%%%%%%%%%%%%%%%%%%%%%%%%%%%%%%%%%%%%%%%%%%%%%%
\section{General structure of kernel contribution}
%%%%%%%%%%%%%%%%%%%%%%%%%%%%%%%%%%%%%%%%%%%%%%%%%%%%%%%%%%%%%%%%%%%
\label{sec:kernel_struct}

Having in mind that we want to investigate the mechanism of
the {\em independence} of evolution kernels on the choice of
the variable $s$ used to enclose phase space,
let us look more closely into the phase space integral
of kernel contribution using $k_\perp$-ordering and $a$-ordering.

General structure of kernel contribution is given by
equation~\eqref{eq:gamma}.
The presence of the residue ${\rm Res}_0$ ensures that only part
proportional to single pole $\frac{1}{\epsilon}$ contributes --
this has to be kept in mind.
For $k_{\perp}$-ordering 
the distribution $\rho(k_1,k_2)$ has general form:
\begin{equation}
\rho(k_1,k_2) = C g^4 \frac{1}{q^4(k_1,k_2)} T(k_1,k_2),
\end{equation}
where $C$ is the color factor specific for each diagram,
$g$ is related to strong coupling by
$g^2=2(2\pi)\alpha_S$, $T(k_1,k_2)$ is dimensionless function
and
$q^2 = \frac{1-\alpha_2}{\alpha_1}\bk_{1\perp}^2 +
       \frac{1-\alpha_1}{\alpha_2}\bk_{2\perp}^2 +
       2\bk_{1\perp}\bk_{2\perp}$.
For $a$-ordering we have:
\begin{equation}
\rho(a_1,a_2) = C g^4 \frac{1}{\alpha_1^2\alpha_2^2}
                \frac{1}{\tilde{q}^4(k_1,k_2)} \tilde{T}(a_1,a_2),
\end{equation}
where $\tilde{T}(a_1,a_2)$ is dimensionless and
$\tilde{q}^2 = \frac{1-\alpha_2}{\alpha_2}\ba_1^2 +
               \frac{1-\alpha_1}{\alpha_1}\ba_2^2 +
               2\ba_1\ba_2$.
The above specific form of $\rho$ enables
immediate factorization of one
$\epsilon$ pole due to integration over
the overall scale variable $\tilde{Q}$,
which we explicitly introduce by means of the identity
$1\equiv \int_0^Q d\tilde{Q} \delta(\tilde{Q}=s(k_1,k_2))$.
The remaining integral is parametrized
using dimensionless variables
$y_i^{\prime}=\frac{k_{i\perp}}{\tilde{Q}}$ or $y_i=\frac{a_i}{\tilde{Q}}$:
\begin{equation}
\label{eq:eps_kT}
\begin{split}
&P^{k_{\perp}}(x) = {\rm Res}_0\Bigg\{
        \left(\frac{\alpha_S}{2\pi}\right)^2
        C\frac{\Omega_{1+2\epsilon}}{(2\pi)^{2+4\epsilon}}
        \int\frac{d\alpha_1}{\alpha_1}\frac{d\alpha_2}{\alpha_2}
\\&~~~~~~\times \delta(1-x-\alpha_1-\alpha_2)
        \int d\Omega_{1+2\epsilon}^{(12)}
         \frac{1}{\mu^{4\epsilon}}
                \int_0^Q d\tilde{Q} \tilde{Q}^{4\epsilon-1}
\\&~~~~~~\times \int_0^1 dy_1^{\prime} dy_2^{\prime}
                (y_1^{\prime}y_2^{\prime})^{1+2\epsilon}\;
                \frac{T(y_1^{\prime},y_2^{\prime},\theta)}
                {q^4(y_1^{\prime},y_2^{\prime},\theta)}\;
                \delta(1-\max\{y_1^{\prime},y_2^{\prime}\})
\bigg\},
\end{split}
\end{equation}
and
\begin{equation}
\label{eq:eps_rap}
\begin{split}
&P^a(x) = {\rm Res}_0\Bigg\{
            \left(\frac{\alpha_S}{2\pi}\right)^2
            C\frac{\Omega_{1+2\epsilon}}{(2\pi)^{2+4\epsilon}}
            \int \frac{d\alpha_1}{\alpha_1}\frac{d\alpha_2}{\alpha_2}
            (\alpha_1\alpha_2)^{2\epsilon}
\\&~~~~~~\times 
            \delta(1-x-\alpha_1-\alpha_2)
            \int d\Omega_{1+2\epsilon}^{(12)}
            \frac{1}{\mu^{4\epsilon}}
            \int_0^Q d\tilde{Q} \tilde{Q}^{4\epsilon-1}
\\&~~~~~~\times 
     \int_0^1 dy_1 dy_2 (y_1y_2)^{1+2\epsilon}\;
                \frac{\tilde{T}(y_1,y_2,\theta)}
                {\tilde{q}^4(y_1,y_2,\theta)}\;
                \delta(1-\max\{y_1,y_2\})
\Bigg\}.
\end{split}
\end{equation}
Now in eqs.~(\ref{eq:eps_kT}) and (\ref{eq:eps_rap})
the pole $\frac{1}{\epsilon}$ 
gets explicitly factorized off in form of the integral 
$\int_0^Q d\tilde{Q}\tilde{Q}^{4\epsilon-1} =
\frac{Q^{4\epsilon}}{4\epsilon}$
and it is now transparent that the integrals of the above equations
feature {\em at least} single $\frac{1}{\epsilon}$ pole.

Possible additional $\frac{1}{\epsilon}$ poles may arise from
internal singularities of the integrands of Feynman diagrams.
They are always connected with integrations over
transverse degrees of freedom ($y_i$).
The longitudinal components can also lead to infra-red (IR)
singularities, when $\alpha_i\to 0$ but this type
of singularities do not lead to additional $\epsilon$-poles, because
in the CFP scheme they are regularized in a non-dimensional
manner~\footnote{%
  For regularization of IR singularities CFP use
  principal value prescription:
  $\frac{1}{\alpha}\rightarrow\frac{\alpha}{\alpha^2+\delta^2}$.
  We also use the following notation of CFP for divergent integrals:
  $\int_0^1d\alpha\frac{\alpha}{\alpha^2+\delta^2}\equiv I_0$ and
  $\int_0^1d\alpha\ln\alpha\frac{\alpha}{\alpha^2+\delta^2}\equiv I_1$
}.

Furthermore, equations~\eqref{eq:eps_kT} and~\eqref{eq:eps_rap}
show explicitly the differences between exclusive kernel
contributions (integrands). This means that exclusive
$MS$ evolution kernels do depend on evolution
time variable.

Equations~\eqref{eq:eps_kT} and~\eqref{eq:eps_rap} are the
starting point for the investigation of the dependence of
inclusive evolution kernels on the upper phase space
limit/evolution time variable.
There will be at least two cases to be considered:
(i) with no additional internal singularities present,
hence terms originating form the expansion of
$(\alpha_1\alpha_2)^{2\epsilon}=1+2\epsilon\ln(\alpha_1\alpha_2)$
can be neglected,
(ii) with the additional $\epsilon$-poles due to internal singularities present,
hence the expansion term $2\epsilon\ln(\alpha_1\alpha_2)$ is contributing.

%%%%%%%%%%%%%%%%%%%%%%%%%%%%%%%%%%%%%%%%%%%%%%%%%%%%%%%%%%%%%%%%%%%
%%%%%%%%%%%%%%%%%%%%%%%%%%%%%%%%%%%%%%%%%%%%%%%%%%%%%%%%%%%%%%%%%%%
\section{Kernels independence on the evolution time variable}
%%%%%%%%%%%%%%%%%%%%%%%%%%%%%%%%%%%%%%%%%%%%%%%%%%%%%%%%%%%%%%%%%%%
\label{sec:mechanisms}

In this section we shall comment on three mechanisms which 
in the CFP factorization scheme
actually protect the {\em independence} of the inclusive NLO DGLAP kernels
of the way the phase space is closed from the above.
It will be demonstrated using example Feynman diagrams.

%%%%%%%%%%%%%%%%%%%%%%%%%%%%%%%%%%%%%%%%%%%%%%%%%%%%%%%%%%%%%%%%%%%
\subsection{Case 1 - no internal singularities}
%%%%%%%%%%%%%%%%%%%%%%%%%%%%%%%%%%%%%%%%%%%%%%%%%%%%%%%%%%%%%%%%%%%
\label{sec:Bx}

\begin{figure}[h!]
  \begin{center}
    \includegraphics[height=2cm]{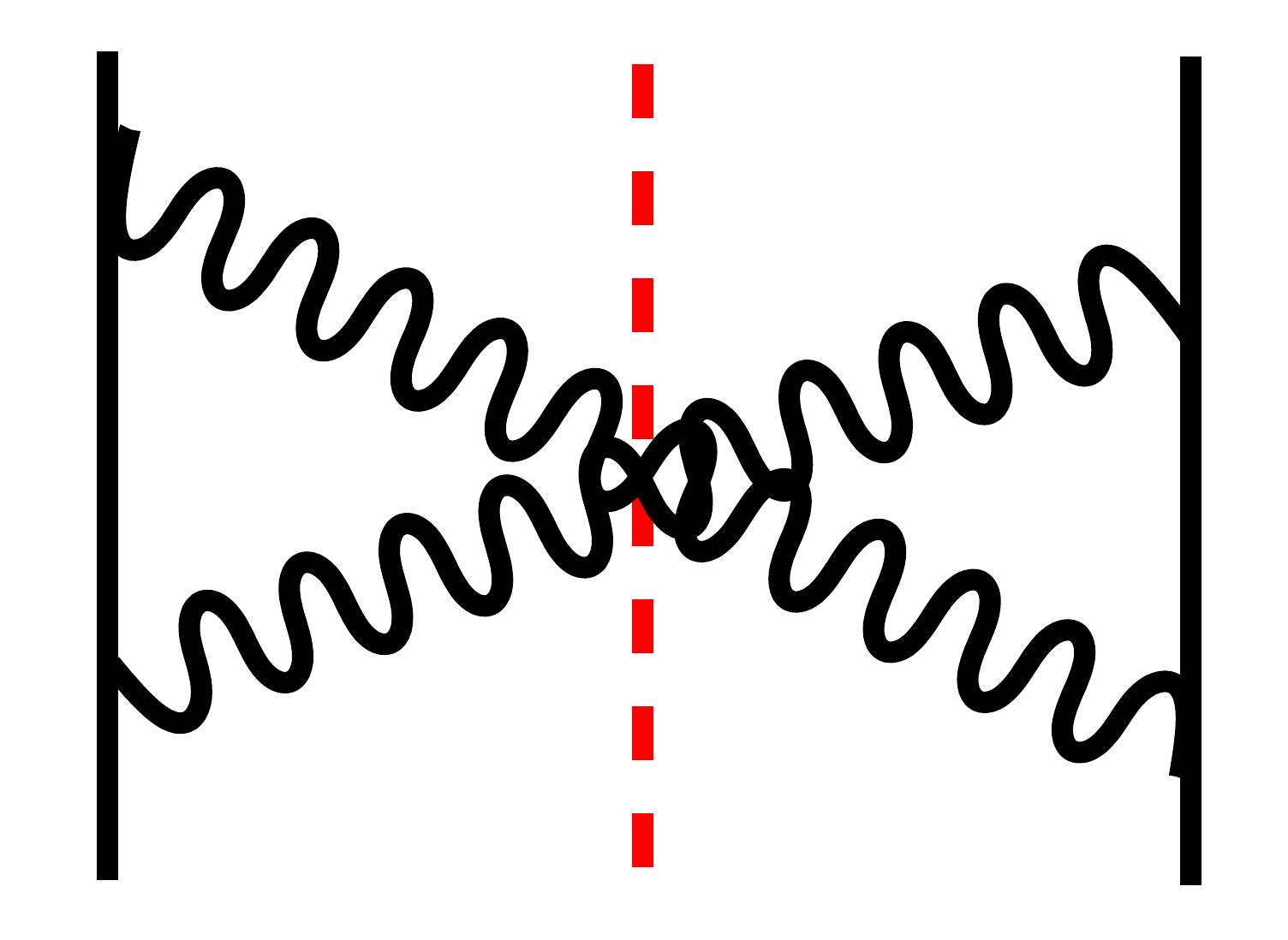}
  \end{center}
\caption{``Crossed ladder graph'', free from any internal singularities.}
\label{fig:Bx}
\end{figure}

Here, the independence will be demonstared
using the example interference diagram of figure~\ref{fig:Bx}.
Starting with the expression of equation~\eqref{eq:eps_rap}
the calculations can be carried out in 4 dimensions
\begin{equation}
\begin{split}
&P_{Bx}^a(x) = N \int
               \frac{d\alpha_1}{\alpha_1}\frac{d\alpha_2}{\alpha_2}
               \delta(1-x-\alpha_1-\alpha_2)\int_0^{2\pi} d\phi\\
&\;\;\;\;\times \int dy_1dy_2 \;y_1y_2
                \frac{\tilde{T}(y_1,y_2,\phi)}
                {\tilde{q}^4(y_1,y_2,\phi)}
                \delta\left(1-\max\{y_1,y_2\}\right),
\end{split}
\end{equation}
where $N$ is normalization constant.
In massless QCD the
integrand has a nice property that $y_1$ and $y_2$
integrals can be combined together into one integral over the
whole space~\footnote{%
  It results from the fact that
  $\tilde{T}$ or $T$ depend on the ratio $y_1/y_2$ only, i.e.
  $\tilde{T}(y_1,y_2,\phi)=\tilde{T}(\lambda y_1,\lambda y_2,\phi)$.%
}
by means of a simple change of variables $y_1=y, y_2=1/y$,
then:
\begin{equation}
\begin{split}
&P_{Bx}^a(x) = N \int\frac{d\alpha_1}{\alpha_1}
               \frac{d\alpha_2}{\alpha_2}
               \delta(1-x-\alpha_1-\alpha_2) \int_0^{2\pi} d\phi
               \int_0^{\infty} dy\;y
               \frac{\tilde{T}(y,1,\phi)}{\tilde{q}^4(y,1,\phi)}.
\end{split}
\end{equation}
Since the change of the phase space enclosure
from {\em angular scale} to {\em transverse momentum} translates
into linear change of variables,
$k_{i\perp}=\alpha_ia_i$, $y_i^{\prime}=\alpha_iy_i$,
and now the $y$ integral extends from zero to $\infty$,
hence the joint integral is manifestly the same
for both kinds of phase space enclosure.

The argument presented in the above example holds for all
kernel contributions free from internal singularities.

%%%%%%%%%%%%%%%%%%%%%%%%%%%%%%%%%%%%%%%%%%%%%%%%%%%%%%%%%%%%%%%%%%%
\subsection{Case 2 - diagram with internal singularity minus counterterm}
%%%%%%%%%%%%%%%%%%%%%%%%%%%%%%%%%%%%%%%%%%%%%%%%%%%%%%%%%%%%%%%%%%%

\begin{figure}[h!]
  \begin{center}
    \includegraphics[height=2cm]{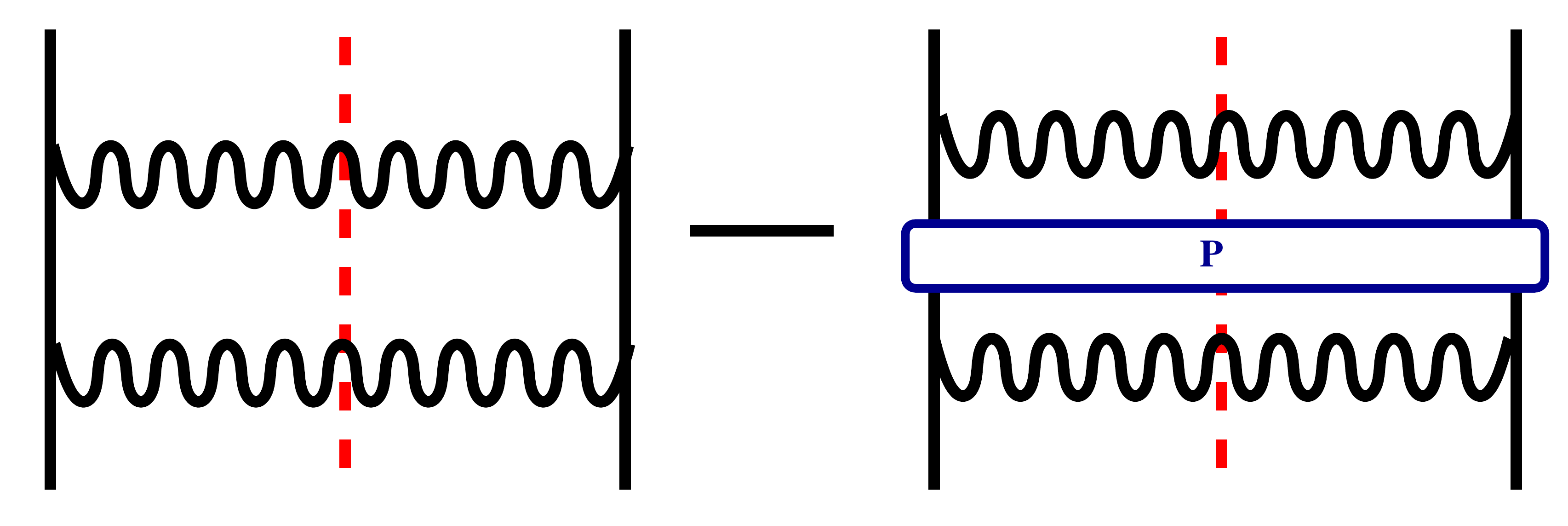}
  \end{center}
\caption{{\sf Bremsstrahlung} graph accompanied by 
  its {\em soft counterterm},
  both featuring double $\epsilon$-poles.}
\label{fig:Br-K}
\end{figure}

Second case is represented by the
{\em double bremsstrahlung} diagram, see figure~\ref{fig:Br-K}.
It has an internal singularity when one of the emitted
gluons is collinear (the other one kept non-collinear).
The additional contribution to the residue due to the
additional (double collinear singularity) terms is of a type
$2\epsilon\ln(\alpha_1\alpha_2)\times\frac{1}{\epsilon^2}$.
This diagram is special because it is accompanied by the
{\em soft counterterm}, which is simply a square of
leading-order (LO) diagram. The soft counterterm is present
due to the factorization scheme (by construction), see
ref.~\cite{Jadach:2009gm, Curci:1980uw, Skrzypek:2009jk}.
We will show, that in this case, the independence of
inclusive kernel contribution on the upper phase space
limit/future evolution time variable is assured by the
presence of the counterterm, which will cancel the additional
$\sim\ln(\alpha_1\alpha_2)$ term.

Since we have shown in section~\ref{sec:Bx} that only terms
leading to additional $\epsilon$ poles can lead to
differences between the two choices of evolution time
variable now we will concentrate only on them. The singular
contribution of the double bremsstrahlung diagram is of a form:
\begin{equation}
\begin{split}
&P_{Br}^{a \;sing}(x) = {\rm Res}_0\Bigg\{
              \frac{N}{\epsilon} \int
                   \frac{d\alpha_1}{\alpha_1}
                   \frac{d\alpha_2}{\alpha_2}
                   \delta(1-x-\alpha_1-\alpha_2)
                   (\alpha_1\alpha_2)^{2\epsilon}
       \int d\Omega_{1+2\epsilon}^{(12)}
\\&~~~~~\times 
              \int_0^1 dy_1dy_2 (y_1y_2)^{1+2\epsilon}
              \frac{1}{\tilde{q}^4(y_1,y_2,\theta)}
              \tilde{T}_2\frac{y_2^2}{y_1^2}
              \delta\left(1-\max\{y_1,y_2\}\right)
\Bigg\}.
\end{split}
\end{equation}
Now combining of the two phase space integrals is not possible any more%
\footnote{%
  In $n=4$ gluing two $y_i$-integrals is still possible
  using the cut-off regularization.
  However, one has to watch out for the cut-off dependent integration's limits.
}
due to the presence of the term $(y_1y_2)^{2\epsilon}$
(coming from phase space) and regularizing $1/y_1^2$ singularity.
There will be differences between $k_{\perp}$ and $a$
parametrizations. There are two sources of this differences.
The first one is simply the difference between the integrals
in both parametrizations. The second is the mixing of double
pole and $\epsilon$ terms from the phase space factor
$(\alpha_1\alpha_2)^{2\epsilon}$.
The difference between the two phase space enclosures (parametrizations) is
\begin{equation}
\begin{split}
&P_{Br}^{a-k_{\perp}}(x) = {\rm Res}_0\Bigg\{
            \frac{C_F^2}{\epsilon}
            \left(\frac{\alpha_S}{2\pi}\right)^2 \int
            \frac{d\alpha_1}{\alpha_1}\frac{d\alpha_2}{\alpha_2}
            \delta(1-x-\alpha_1-\alpha_2)
            2\tilde{T}_2
\\&\;\;\;\times \bigg[ \ln\left(\frac{\alpha_2}{\alpha_1}\right) -
              \ln\left(\frac{\alpha_1}{\alpha_2}\right) +
              2\epsilon\ln(\alpha_1\alpha_2)
              \frac{1}{\epsilon} \bigg]
\Bigg\},
\end{split}
\label{eq:Br_a-kT}
\end{equation}
where coefficient 
$\tilde{T}_2 = \frac{1}{4}(1+(1-\alpha_1)^2)(1+x^2/(1-\alpha_1)^2)$ 
comes from the product of the numerators of two LO kernels.

For the counterterms, which are much simpler, due to their LO structure,
the difference between integrals in $k_{\perp}$
and $a$ space is only due to the phase space factor $\alpha_2^{2\epsilon}$:
\begin{equation}
\begin{split}
&P_{ct}^{a-k_{\perp}}(x) = {\rm Res}_0\Bigg\{
	  C_F^2
          \left(\frac{\alpha_S}{2\pi}\right)^2
          \int d\alpha_1d\alpha_2 \delta(1-x-\alpha_1-\alpha_2)
          2\ln(\alpha_2)
          \frac{2 \tilde{T}_2}{\alpha_1\alpha_2}
\Bigg\}.
\end{split}
\label{eq:ct_a-kT}
\end{equation}
It is almost manifest that the integrals of eqs.~\eqref{eq:Br_a-kT}
and~\eqref{eq:ct_a-kT} are the same, which means that
\begin{equation}
P_{Br}^{a-k_{\perp}}(x) = P_{ct}^{a-k_{\perp}}(x).
\end{equation}
Summarizing, there is an exact cancellation of differences between the
two choices of upper phase space limit among the
bremsstrahlung diagrams and their soft counterterms.
It results in the independence of the kernel contribution
on the two choices of the phase space enclosure variable under consideration.
On general ground, the same statement should hold for other choices, 
for example for the total virtuality.
However, in case of virtuality it is much more difficult
to show, without actually performing the integration,
that the final result for the inclusive NLO kernel is the same
as in the above cases of $k_\perp$-ordering or $a$-ordering.

We want to emphasize the crucial role of $MS$-like
terms ($(\alpha_1\alpha_2)^{2\epsilon}$ and $\alpha_2^{2\epsilon}$)
in restoring the independence of the kernel contribution on
the choice of phase space encloser, which can be seen explicitly
from equations~\eqref{eq:Br_a-kT} and~\eqref{eq:ct_a-kT}.

In the above discussion we were analyzing certain contributions from
double bremsstrahlung diagrams and soft counterterms as
representing the difference between the cases of phase space
enclosure using maximum $k_\perp$ or, alternatively, maximum
angular scale $a$.
In fact, these terms are {\em completely absent} in case of
maximum $k_\perp$, which means that the maximum $k_\perp$ is
effectively representing formal scale parameter $\mu$ of the
dimensional regularization $MS$. These terms are also nonzero
for other popular choices of the phase space enclosure like
maximum $k$-minus and total virtuality.

%%%%%%%%%%%%%%%%%%%%%%%%%%%%%%%%%%%%%%%%%%%%%%%%%%%%%%%%%%%%%%%%%%%
\subsection{Case 3 -  two-real-gluon internal singularity versus virtual diagrams}
%%%%%%%%%%%%%%%%%%%%%%%%%%%%%%%%%%%%%%%%%%%%%%%%%%%%%%%%%%%%%%%%%%%

\begin{figure}[h!]
  \begin{center}
    \includegraphics[height=2cm]{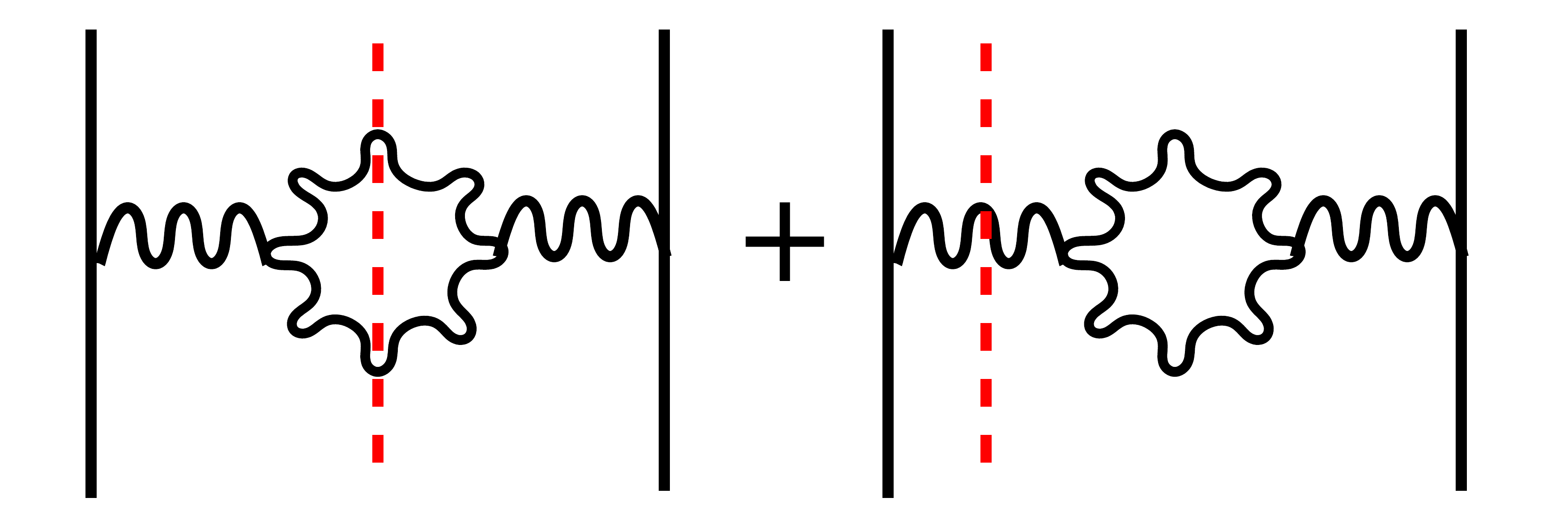}
  \end{center}
\caption{Gluon pair production diagram and the corresponding
virtual diagram (vacuum polarization).}
\label{fig:Vg}
\end{figure}

The third case is a class of diagrams with an {\em internal
singularity} due to parton pair emission
from the ladder, for which the independence of the evolution
time variable in assured by the corresponding virtual
diagrams. The example diagram of this type is shown in
Figure~\ref{fig:Vg}.
This is a diagram with gluon pair production, where the internal
singularity occurs when the invariant mass of the produced
pair goes to zero.
The additional $\epsilon$ pole originating from
the singularity $1/k^2=1/(k_1+k_2)^2$, together with
$(\alpha_1\alpha_2)^{2\epsilon}$ factor, will lead to
a familiar mixing terms in the residue.
This mixing terms leads to the differences between
real parton integrals once the
different choices of the upper phase space enclosure are applied.
Of course, in CFP scheme there is a mechanism
which brings back the independence of the inclusive kernels on that.
In this case the contributions
of the corresponding divergent virtual diagram do this job.

In this case the calculations are technically more complicated
and have to remain beyond the scope of this contribution.
In fact the independence was explicitly checked by means of switching
from the {\em angular scale} to the
{\em overall virtuality} as the upper phase space limiting variable
as they are best suited for the singularity structure of these diagrams.

%%%%%%%%%%%%%%%%%%%%%%%%%%%%%%%%%%%%%%%%%%%%%%%%%%%%%%%%%%%%%%%%%%%
%%%%%%%%%%%%%%%%%%%%%%%%%%%%%%%%%%%%%%%%%%%%%%%%%%%%%%%%%%%%%%%%%%%
\section{Conclusions}
%%%%%%%%%%%%%%%%%%%%%%%%%%%%%%%%%%%%%%%%%%%%%%%%%%%%%%%%%%%%%%%%%%%

We investigate the mechanism which ensures
the independence of the NLO DGLAP evolution kernels
calculated within Curci-Furmanski-Petronzio scheme
on the choice of the upper phase space limiting variable
$s(k_1,k_2)$.

It was shown that for different groups of Feynman diagrams
there are  three mechanisms which work in order
to compensate the differences due to change of the type of $s(k_1,k_2)$.
The independence is demonstrated explicitly
in case of transverse momentum and rapidity related variable $a$.
(The investigation has been carried out also for different choices
like overall virtuality $q^2$, maximum light-cone variable $k$-minus,
but no details are reported here.)

We have show that the mechanisms protecting this property involves
either soft counterterms of CFP scheme or virtual diagrams.

In case of the MC implementation of the exclusive PDFs,
see refs.~\cite{Jadach:2009gm, Jadach:2010ew, Skrzypek:2009jk},
keeping track of these phenomena in the kernel calculations
is useful for understanding what happens
while switching from one version
of the evolution time variable in the Monte Carlo to the other,
more details will be provided in ref.~\cite{ifjpan-2009-06}.

\vspace{4mm}
\noindent
{\bf Acknowledgments}\\
We would  like to acknowledge support and warm hospitality of CERN EP/TH  (S.J. and M.S.), University of Manchester (A.K.) and IPPP Durham (M.S.) during the preparation of this work.

%%%%%%%%%%%%%%%%%%%%%%%%%%%%%%%%%%%%%%%%%%%%%%%%%%%%%%%%%%%%%%%%%%%%%%%%%%%%
% % \begingroup\begin{thebibliography}{10}
% \bibliographystyle{utphys_spires}
% % \bibliographystyle{h-physrev3}
% \bibliography{radcor}
% % \end{thebibliography}\endgroup

\providecommand{\href}[2]{#2}

%%%%%%%%%%%%%%%%%%%%%%%%%%%%%%%%%%%%%%%%%%%%%%%%%%%%%%%%%%%%%%%%%%%%%%%%%%%

\end{document}